\documentclass[]{emulateapj}
\usepackage{apjfonts}
\newcommand{\be}{\begin{equation}}
\newcommand{\ee}{\end{equation}}
\newcommand{\bea}{\begin{eqnarray}}
\newcommand{\eea}{\end{eqnarray}}
\newcommand{\Msun}{M_{\odot}}

\def\Mhalo{M_{\rm halo}}
\def\fesc{f_{\rm esc}}
\def\frun{f_{\rm run}}
\def\vrun{V_{\rm run}}
\def\mmax{M_{\rm max}}
\def\Msun{\ M_\Sol}

\def\kms{\ {\rm km\, s}^{-1}}
\def\cmt{\ {\rm cm}^{-2}}
\def\cm3{\ {\rm cm}^{-3}}

\shortauthors{CONROY \& KRATTER}
\shorttitle{Runaway Stars and the Escape of Ionizing Radiation from
  High-Redshift Galaxies}

\begin{document}
\journalinfo{The Astrophysical Journal}

\title{Runaway Stars and the Escape of Ionizing Radiation from
  High-Redshift Galaxies}

\author{Charlie Conroy \& Kaitlin M. Kratter}
\affil{Harvard-Smithsonian Center for Astrophysics, Cambridge, MA,
  USA}

\begin{abstract}

  Approximately 30\% of all massive stars in the Galaxy are runaways
  with velocities exceeding $30\kms$.  Their high speeds allow them to
  travel $\sim0.1-1$ kpc away from their birth place before they
  explode at the end of their several Myr lifetimes.  At high
  redshift, when galaxies were much smaller than in the local
  universe, runaways could venture far from the dense inner regions of
  their host galaxies.  From these large radii, and therefore low
  column densities, much of their ionizing radiation is able to escape
  into the intergalactic medium.  Runaways may therefore significantly
  enhance the overall escape fraction of ionizing radiation, $\fesc$,
  from small galaxies at high redshift.  We present simple models of
  the high-redshift runaway population and its impact on $\fesc$ as a
  function of halo mass, size, and redshift.  We find that the
  inclusion of runaways enhances $\fesc$ by factors of $\approx1.1-8$,
  depending on halo mass, galaxy geometry, and the mechanism of
  runaway production, implying that runaways may contribute $50-90\%$
  of the total ionizing radiation escaping from high-redshift
  galaxies.  Runaways may therefore play an important role in
  reionizing the universe.
  
\end{abstract}



\section{Introduction}
\label{s:intro}

Observations of the cosmic microwave background \citep{Dunkley09,
  WMAP7} and the absorption spectra of high-redshift quasars
\citep{Fan06} imply that the universe was reionized sometime between
$6<z<12$.  The rapid drop in the quasar luminosity function at high
redshift suggests that massive stars provide the bulk of the ionizing
luminosity responsible for reionization \citep{Madau99, Loeb01,
  Bolton05}.  Stars at high redshift are born within dense
proto-galactic fragments, and thus much of their ionizing luminosity
is thought to be attenuated by neutral hydrogen.  The fraction of
their ionizing luminosity that escapes into the intergalactic medium
(IGM) and is available to reionize the universe, $\fesc$, is a key
quantity in determining how reionization occurred.

Models for reionization that reproduce the observed UV luminosity
function of high-redshift galaxies require both $\fesc\gtrsim0.1$ and
significant contributions from very faint galaxies
\citep[e.g.,][]{Pawlik09, Srbinovsky10, Haardt12, Kuhlen12,
  Bouwens12}.  These requirements place very demanding constraints on
models of high-redshift galaxies.

In the local universe, observations of both normal star forming and
starburst galaxies favor escape fractions in the range of a few
percent \citep{Leitherer95, Bland-Hawthorn99, Heckman01, Grimes07}.
The escape fraction seems to be similarly low at $z\sim1$
\citep{Siana07, Cowie09, Siana10}.  At higher redshifts ($z\sim3$),
observations suggest that $\fesc$ may increase to $\sim10\%$, at least
for the relatively massive, highly star-forming galaxies for which
measurements of $\fesc$ have been made \citep{Steidel01, Giallongo02,
  Shapley06, Inoue06, Siana07, Iwata09, Siana10}.  However,
constraints on $\fesc$ at $z>2$ from gamma-ray burst afterglows favor
low values of $\fesc\approx0.02$ \citep{Chen07}.  While the data
suggest that escape fractions may increase out to $z\sim3$, it is
worth noting that direct measurements of ionizing photons during the
epoch of reionization are impossible, owing to the high opacity of the
IGM.  Our understanding of how reionization proceeded therefore
depends crucially on models and simulations that follow the escape of
ionizing photons.

Simple theoretical estimates of $\fesc$ can be obtained from analytic
models of galaxies and HII regions \citep{Dove94, Haiman97, Madau99,
  Dove00, Ricotti00, Wood00, Fernandez11}.  These models produce
estimates of $\fesc$ that vary widely from $10^{-3}$ to $\sim1$
depending on dark matter halo mass, clumpiness of the interstellar
medium (ISM), gas density profile, and redshift.  Sophisticated
hydrodynamical simulations also predict a wide range of escape
fractions \citep{Razoumov06, Razoumov07, Gnedin08, Wise09, Razoumov10,
  Yajima11}.  

While some of the differences between model predictions can be
attributed to different choices for physical parameters such as halo
mass and redshift, numerical methods and sub-grid models also
contribute to the quoted discrepancies.  For example, the geometry of
the ISM and the relative distribution of stars and gas has a strong
influence of $\fesc$, and both are controlled by uncertain
prescriptions for star formation and supernovae feedback.
\citet{Clarke02} argued that there is a critical star formation rate
above which the porosity of the ISM is approximately unity. A highly
porous ISM is expected to result in a high escape fraction
\citep{Fujita03}.  Indeed, simulations are able to achieve high escape
fractions only when stellar feedback is very strong and hence the
porosity is very high.  \citet{Ricotti02a} has considered globular
clusters as efficient sources of ionizing photons due to their high
star formation efficiencies and high porosities.  In this paper we
consider an alternative scenario that is capable of producing very
high escape fractions without appealing to a highly porous ISM: the
contribution from runaway stars.

A significant fraction of massive stars in the Galaxy are moving at
high velocity ($>30\kms$) and are therefore known as runaways
\citep[e.g.,][]{Blaauw61, Gies86}.  Two formation channels have been
proposed for the formation of runaways: 1) via dynamical ejections
from young dense stellar systems \citep{Gies86, Fujii11, Perets12},
and 2) due to the explosion of a companion star \citep{Zwicky57,
  Blaauw61, Gott70, PortegiesZwart00, Eldridge11}.  A combination of
these two scenarios has also been considered
\citep{Pflamm-Altenburg10}. Tracing the orbits of runaways back in
time has provided evidence that both channels contribute to the
runaway population \citep{Hoogerwerf01}.  The available data on
runaways allow for a wide range in the fraction of massive stars that
are runaways, $\frun$, from $0.1-0.5$ depending on definition and
correction for completeness and observational biases
\citep[e.g.,][]{Gies86, Gies87, Stone91, Tetzlaff11}.

The goal of this paper is to assess the effect of runaways on the
escape of ionizing radiation from high-redshift galaxies.
Qualitatively we expect them to be important when galaxy sizes are
smaller than $\sim100$ pc, as then even relatively slow runaways with
short main sequence lifetimes can travel from a galaxy's center to its
outskirts before exploding.  In this paper we will explore the effect
of runaways by building simple analytic models of high-redshift
galaxies and modeling runaways formed through both the dynamical and
supernova mechanisms. The potential importance of runaway stars on the
escape of ionizing photons was first noted by \citet{Dove94} in the
context of the Galaxy.

Throughout this paper we adopt cosmological parameters consistent with
the 7th year {\it WMAP} estimates \citep{WMAP7}, namely
$\Omega_\Lambda=0.73$ and $\Omega_m=0.27$, and a Hubble constant
of $H_0=70$ km s$^{-1}$ Mpc$^{-1}$.

\section{Model}

Our goal is to construct simple models for the escape fraction of
ionizing radiation from high-redshift galaxies both with and without
the contribution of runaways.  Our basic setup is motivated by and
follows the approach of \citet[][]{Ricotti00}, and references therein.

\subsection{Galaxy Models}

We begin by considering the relation between cold dark matter halo
mass and virial radius derived from cosmological $N-$body simulations
\citep{NFW97}:
\noindent
\be
\label{e:r}
r_{\rm vir} = 0.96\,\ {\rm kpc}\, \bigg(\frac{M_{\rm halo}}{10^8\,\Msun}\bigg)^{1/3} \bigg(
\frac{\Omega_m}{\Omega_m(z)}\frac{\Delta_c}{200}\bigg)^{-1/3} \bigg( \frac{1+z}{10}\bigg)^{-1},
\ee
\noindent
where $\Omega_m$ and $\Omega_m(z)$ are the matter densities of the
universe at the present epoch and redshift $z$, respectively, in units
of the critical density ($\rho_{\rm crit}=3H_0^2/8/\pi G$), and
$\Delta_c$ is the overdensity threshold of virialized dark matter
halos, which depends on $\Omega_m(z)$ \citep{Bryan98}.  For our fiducial
model we adopt $z=10$.

We will assume that 100\% of the baryons are in the cold gas phase
\citep[i.e., the stellar fraction is $\approx0$ at these early epochs;
see e.g.,][]{Ricotti02b, Wise09}, implying:
\noindent
\be
\label{e:m}
M_{\rm gas} = f_b\, M_{\rm halo},
\ee
where $f_b=0.17$ is the universal baryon fraction \citep{WMAP7}.
The relation between the gas scale radius and the halo virial radius
is taken to be \citep{Mo98}:
\noindent
\be
r_g = \frac{j_d\lambda}{\sqrt{2}f_b}\, r_{\rm vir},
\ee
\noindent
where $\lambda$ is the dimensionless spin parameter, which for halos
in simulations has a mean value of 0.04 \citep{Bullock01}.  The galaxy
angular momentum is a fraction $j_d$ of that of the halo.  We assume
$j_d/f_b=1$ and $\lambda=0.05$, as these parameters provide a good fit
to the observed size distribution of local galactic disks
\citep{Mo98}.

High-resolution cosmological hydrodynamic simulations produce a wide
range of galaxy morphologies at high redshift, ranging from spherical
to well-ordered disks \citep{Wise09,Pawlik11,Romano-Diaz11,Wise12}.
Because the true nature of these galaxies remains unknown, and at
present unobservable, we consider two galaxy geometries, spherical and
disk-like.  In our spherical model the gas has an exponential number
density profile, $n(r) = n_0\, e^{-r/r_g}$.  In our disk model, we
consider a gas profile with $n(r,z)=n_0\,e^{-R/r_g}e^{-z^2/z_g^2}$,
where $z_g$ is the disk scale height.  We consider models where the
disk height-to-disk scale length ratio varies with mass as:
\noindent
\be
\frac{z_g}{r_g} = 0.2 \bigg(\frac{\Mhalo}{10^8\Msun}\bigg)^{-2/3}.
\ee
\noindent
The mass dependence of $z_g/r_g$ follows from the assumption of an
isothermal disk \citep[e.g.,][]{Wood00}. To ensure that $z_g/r_g$
never exceeds unity we sum its inverse in quadrature with one:
$(z_g/r_g)^{-2} \rightarrow (z_g/r_g)^{-2}+1$.

In our fiducial model massive stars and stellar clusters are born
tracing the gas density profile.  We will also consider a model where
the stellar density is proportional to the gas density squared
($\rho_\ast\propto\rho_g^2$) as in \citet{Wood00}. The slope of the
observed Kennicutt-Schmidt relation between star formation rate
surface density and gas surface density falls in between these two
extremes \citep{Kennicutt98}.  The gas and stellar density
distributions are truncated at $5r_g=0.2r_{\rm vir}$ (and $5z_g$ for
disks).  In summary, for our fiducial set of model parameters, the
halo mass uniquely specifies the distribution of gas and stars within
the halo.

\subsection{Photon Escape Fraction}

Our next task is to compute $\fesc$.  As discussed in \citet{Dove94}
and \citet{Ricotti00}, the HII regions created by massive stars will
not be spherical if the local gradient in the gas density profile is
large.  While the HII regions are allowed to be non-spherical in our
model, the equations presented in this section assume that the gas
distribution is spherical; the generalization to disk geometries is
straightforward.  Consider a star a distance $r$ from the center of
the galaxy.  A point residing a distance $r'$ from the star at an
angle $\theta$ from the line connecting the galaxy center and the star
will be at a distance $R=\sqrt{r^2+r'^2+2rr'{\rm cos}\,\theta}$ from
the center of the galaxy \citep[see Figure 1 in][for a sketch of the
geometrical configuration]{Ricotti00}\footnote{There appears to be an
  erroneous substitution of sin for cos in their equation for $R$.} .
At each angle $\theta$ one can compute the quantity:
\noindent
\be
\label{e:ftheta}
f(\theta) = 1-\frac{4\pi\alpha_H^B}{Q}\int_0^\infty n_H(R)^2\,r'^2\,{\rm d}r',
\ee
\noindent
where $Q\equiv Q_H$ is the hydrogen ionizing luminosity, $\alpha_H^B$
is the hydrogen case B recombination coefficient, and $n_H$ is the
number density of hydrogen.  We adopt
$\alpha_H^B=2.6\times10^{-13}\cm3\, {\rm s}^{-1}$, appropriate for gas
at $T=10^4$ K.  Values of $f>0$ indicate a non-zero escape fraction,
while $f\leq0$ indicates that 100\% of the ionizing radiation is
absorbed.  The critical angle for escape, $\theta_c$ is defined
through the equation $f(\theta_c)=0.0$.  Notice that $\theta_c$ is a
function of $Q$ and $r$.  The angle-averaged fraction of ionizing
photons escaping the galaxy is then computed by integrating
$f(\theta)$ over a solid angle:
\noindent
\be
\langle \fesc \rangle = \frac{1}{4\pi}\, \int_0^{\theta_c}\,
f(\theta)\,2\pi\,{\rm sin}\, \theta\, {\rm d}\theta.
\ee
The total escape fraction of the galaxy is then computed by
integrating over $r$ and $Q$:
\noindent
\be
\label{e:fesc}
\fesc = \int\int \langle \fesc\rangle\,P(r)\,P(Q)\, {\rm
  d}\,r{\rm d}Q,
\ee
where $P(r)$ and $P(Q)$ are the probabilities of a star being at
location $r$ and with ionizing luminosity $Q$.

We have neglected the effect of overlapping HII regions in our
analytic framework.  This is a common simplification made in the
modeling of the ionizing escape fraction \citep{Dove94, Ricotti00}.
The effect of including HII overlap would be to increase $\fesc$,
although the low expected star formation rates in $z=10$ halos
\citep{Wise09} suggests that HII overlap may not be significant.

In the absence of runaways, $P(r)$ and $P(Q)$ are straightforward to
specify (our prescription for including runaways is described in
$\S$\ref{s:run}).  The former is determined by the gas density
profile, while the latter is taken to be the observed ionizing
luminosity function (LF) in local galaxies.  The LFs of OB
associations in nearby galaxies have power-law indices that range
between $-1.5$ and $-2.5$ over the range $48\lesssim{\rm
  log}(Q/s^{-1})\lesssim51$ \citep{Kennicutt89, McKee97, Whitmore99,
  vanZee00, Azimlu11}.  \citet{Oey98} argued that the data are
consistent with a universal power-law slope of $-2.0$.  We therefore
adopt an ionizing LF with a $P(Q)\propto Q^{-2.0}$ over the range
$48<{\rm log}(Q/s^{-1})<51$.  For reference, log$(Q)=48$ corresponds to
a single $20\Msun$ star.  We will explore variation in the lower
luminosity cutoff, $Q_{\rm min}$, in $\S$\ref{s:res}.  Notice that our
treatment of $P(Q)$ naturally accounts for the fact that massive stars
are clustered.

Current evidence favors very little if any dust in low mass galaxies
at high redshift, as inferred from their very blue UV colors
\citep{Bouwens12b}. The lack of dust is also consistent with their
expected low metallicities. Our fiducial model is therefore dust-free,
but we nevertheless explore the effect of dust on $\fesc$.  To
incorporate dust into our models, we adopt a wavelength dependent dust
cross section, $\sigma(\lambda)$ from \citet{Pei92}, as updated in
\citet{Gnedin08}.  Essentially nothing is known about the dust
composition nor grain size distribution in high-redshift galaxies, so
we consider both Large and Small Magellanic Cloud-type dust (LMC and
SMC, respectively) to give a sense of the variation expected from
different grain populations.  To make the computations simple, we have
taken averages of $\sigma(\lambda)$ over ionizing photons, weighted by
model stellar flux distributions from \citet{Smith02}.  The average
depends weakly on the effective temperature, varying by $\sim10\%$
over the relevant range.  The resulting flux-weighted mean cross
sections for LMC and SMC dust are $\sigma=1\times10^{-21}\cmt$ and
$\sigma=5\times10^{-22}\cmt$, respectively.  For each $r,\theta$ pair
we can compute $\tau_d=\sigma N_H$, where $N_H$ is the column density
of hydrogen (we adopt the observed dust-to-gas ratios of the LMC and
SMC, which likely overestimates the influence of dust at high
redshift).  The hydrogen ionizing luminosity $Q$ that enters into
Equation \ref{e:ftheta} is then simply attenuated by $e^{-\tau_d}$.

With the gas density profile, ionizing LF, and dust model specified,
$\fesc$ for non-runaway stars is then uniquely determined as a
function of halo mass.  We now consider the addition of runaways.

\subsection{Runaways}
\label{s:run}

For the purposes of computing $\fesc$, runaways differ from
non-runaways in two respects: they travel from their birth
environment, resulting in a change to $P(r)$, and the runaway
population may not reflect the overall massive star population in its
mass distribution and therefore its ionizing LF.  As noted in the
Introduction, there are two mechanisms capable of producing runaways
in the Galaxy: 1) dynamical encounters in young dense stellar systems,
and 2) the explosion of a close companion.  In order to explore the
effects of runaways on $\fesc$ we develop simple models for both
scenarios, motivated by the observed runaway population in the local
universe.

In the dynamical mechanism, runaways inherit a high velocity, $\vrun$,
not long after their birth.  We model runaways produced via this
mechanism by taking a fraction of all massive stars, $\frun$, and
giving them a velocity $\vrun$ at birth with a random direction in the
galaxy.  The runaway is allowed to travel for a time equal to its main
sequence lifetime.  The effect of the potential due to the galaxy and
halo mass on the motion of the runaways is neglected\footnote{The
  circular velocity of dark matter halos at $0.1r_{\rm
    vir}\approx3r_g$ is $<30\kms$ for $\Mhalo<10^9\Msun$ and $<60\kms$
  for $\Mhalo<10^{10}\Msun$ \citep{NFW97}.  For the typical runaway
  velocities we consider ($\vrun>30\kms$), the effect of the halo
  potential on the runaway trajectory can be safely neglected at low
  halo masses, but it may have a modest effect at the highest masses
  we consider.}.

In the simplest version of this model we assume that the runaways are,
at birth, a random sampling of the overall massive star population.
However, in reality one may expect the maximum mass of runaways
produced in this mechanism to be lower than the upper mass of all
stars.  This may occur because dynamical encounters tend to eject the
lowest mass star in the encounter \citep{Anosova86}.  We model this
effect by considering a possible truncation in stellar mass, $\mmax$,
of the runaway mass function.

We explore models where $\vrun$ either takes a single value or a
distribution.  The numerical models of \citet{Fujii11} and
\citet{Perets12} produce runaways with a power-law distribution of
$\vrun$ with an index of $\approx-1$ and $\approx-1.5$, respectively,
over the range $20\kms<\vrun<100\kms$.  Observations of runaways in
the Galaxy seem to favor a Maxwellian velocity distribution with a
dispersion of $\approx30\kms$ \citep{Stone91, Tetzlaff11}, implying a
mean runaway velocity of $\approx45\kms$.  We consider both of these
distributions and their effects on $\fesc$.

For the dynamical model we assume that the runaway LF is equal to the
non-runaway LF up to a maximum ionizing luminosity corresponding to a
mass of $100\Msun$.  Conceptually, this corresponds to replacing
clusters with single stars.  See the discussion in \cite{Oey98}, where
they find that the single-star LF does not significantly differ from
the cluster LF for a range of assumptions.  To calculate
the influence of runaways, we must translate a given luminosity into a
stellar mass so that we may assign stellar lifetimes.  We have
constructed an ionizing luminosity-stellar mass relation, $Q(m)$, by
combining the stellar interior models of \citet{Schaller92} with the
stellar spectral library of \citet{Smith02} at $0.05Z_\Sol$.  We then
use the main-sequence lifetime-mass relation, $t_H(m)$, presented in
\citet{Ekstrom12} for their rotating stellar models.  At fixed mass,
the main sequence lifetime varies by $\pm20\%$ depending on the
modeling of the stellar interiors and metallicity \citep{Marigo01,
  Schaerer02, Ekstrom12}.

The second mechanism we consider is the supernova origin of runaways.
In this scenario the runaway was at some point in a binary with a more
massive companion.  When the companion explodes, the surviving star
inherits some fraction of the pre-explosion orbital velocity.  As in
the previous scenario, we specify the fraction of runaways, $\frun$,
produced in this way.  In addition, we must specify the distribution
of mass ratios, $q\equiv M_2/M_1$, which we assume to be flat over the
range $0.1<q<1.0$, consistent with constraints from the Galaxy
\citep{Sana11}.  We also assume that the primary mass, $M_1$ is drawn
from a \citet{Salpeter55} initial mass function.  Calculation of the
runaway velocity requires knowledge of the period distribution,
initial-final mass relation for the more massive companion, and, for
binaries that remain bound, knowledge of the post-explosion orbital
properties \citep{Gott72}.  Consideration of these factors is beyond
the scope of the present work \citep[see e.g.,][]{PortegiesZwart00,
  Eldridge11}.  Instead, we take an approach that is similar to our
treatment of $\vrun$ in the dynamical model: we simply consider a
fixed value of $\vrun$ that is similar to the observed average runaway
velocity of Galactic O stars.

The supernova mechanism differs from the dynamical mechanism in at
least two important respects.  First, the lifetime of a runaway is
shorter in the supernova scenario because the runaway is formed only
after the death of a more massive companion, while in the dynamical
scenario the velocity kick is imparted at birth. As we will see below,
this effect can be significant.  The ratio in median runaway lifetimes
between the two mechanisms is $\sim3$, so runaways produced via close
dynamical encounters can have a substantially larger impact on
$\fesc$.  In reality, runaways formed in the dynamical channel will be
born some finite time after their birth, tempering this difference.
The second distinction is that the runaway ionizing LFs are different
in the two models. While in the dynamical model runaways have the same
LF as the non-runaways, in the supernova model the runaway LF is
determined by a combination of the IMF, main sequence lifetime-stellar
mass relation, and binary properties.  In practice, the supernova
runaway LF is not very different from the dynamical runaway LF --- it
is approximately a power-law with an index of $\sim-1.7$.

Both of these models are implemented via Monte Carlo simulations.  For
our fiducial model we adopt the dynamical mechanism for runaways with
$\frun=0.3$, $\vrun=60\kms$, and $\mmax=100\Msun$.  In addition to
this fiducial model, a variety of model permutations will be
considered in $\S$\ref{s:res}.

\subsubsection{Runaways at High Redshift}

We have adopted runaway models based on observations of massive stars
in the Galaxy, so it is reasonable to ask whether the population may
be different at high redshift.  For both the dynamical and the
supernova model, the binary population affects the runaway fraction,
with more binaries leading to more runaways. High-mass binary
formation may be an inevitable consequence of disk fragmentation
driven by high accretion rates \citep{Kratter10}.  Indeed, recent
simulations of population III star formation suggest that binaries may
be ubiquitous even in zero metallicity systems \citep{Turk09,
  Greif11}.  It is therefore plausible to assume that the binary
fraction at high redshift is similar to that found in the local
universe.

In the dynamical ejection model the number of runaways depends on star
cluster properties. Following \cite{Fujii11}, who suggest that
clusters produce a fixed number of runaways independent of cluster
mass, one expects the runaway {\it fraction} to scale inversely with
cluster mass. Therefore, if more stars are born in smaller$-N$
clusters, a higher fraction will be runaways. Simple estimates of the
characteristic fragmentation mass, $M_c$, in a $Q=1$ disk
\citep{Toomre64} suggest that it should be smaller at high redshift.
In a gaseous disk $M_c$ scales as $M_c\sim\Sigma\sigma^2/\kappa^2$
where $\Sigma$ is the mass surface density, $\sigma$ is the velocity
dispersion in the gas, and $\kappa$ is the epicyclic frequency.
Adopting Keplerian rotation and using Equations \ref{e:r} and
\ref{e:m} yields $M_c\sim\sigma^3(1+z)^{-3/2}$.  Unless $\sigma$ is
considerably higher in high-redshift disks compared to the Galaxy, we
expect $M_c$ to be lower at high redshift. Everything else being
equal, this would suggest that the runaway fraction may increase with
redshift.  However, the efficiency of producing runaways must decline
for the lowest cluster masses, where the number of high mass stars
that can form is limited by the cluster mass.

Given all these considerations, we believe that adopting a minimal
model, where the runaway fraction does not evolve to high redshift, is
acceptable.

\section{Results}
\label{s:res}

\begin{figure}[!t]
\center
\resizebox{3.4in}{!}{\includegraphics{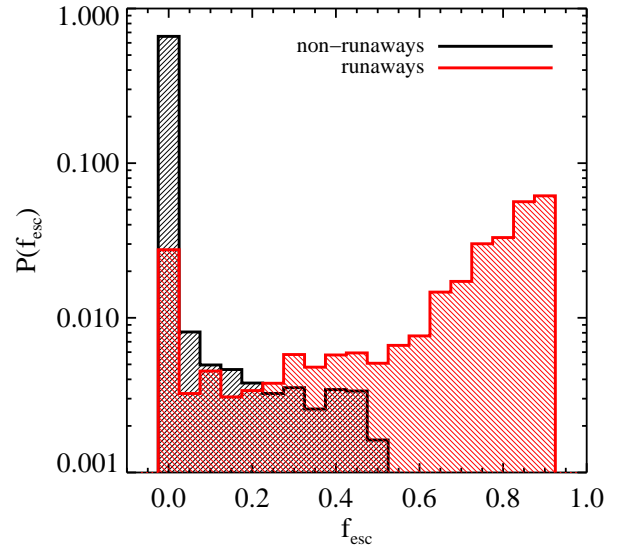}}
\caption{Distribution of ionizing escape fractions for individual
  massive stars in the fiducial model for $\Mhalo=10^8\Msun$.  The
  non-runaways (black) and runaways (red) are shown separately.}
\label{fig:pfesc}
\end{figure}

\begin{figure}[!t]
\center
\resizebox{3.4in}{!}{\includegraphics{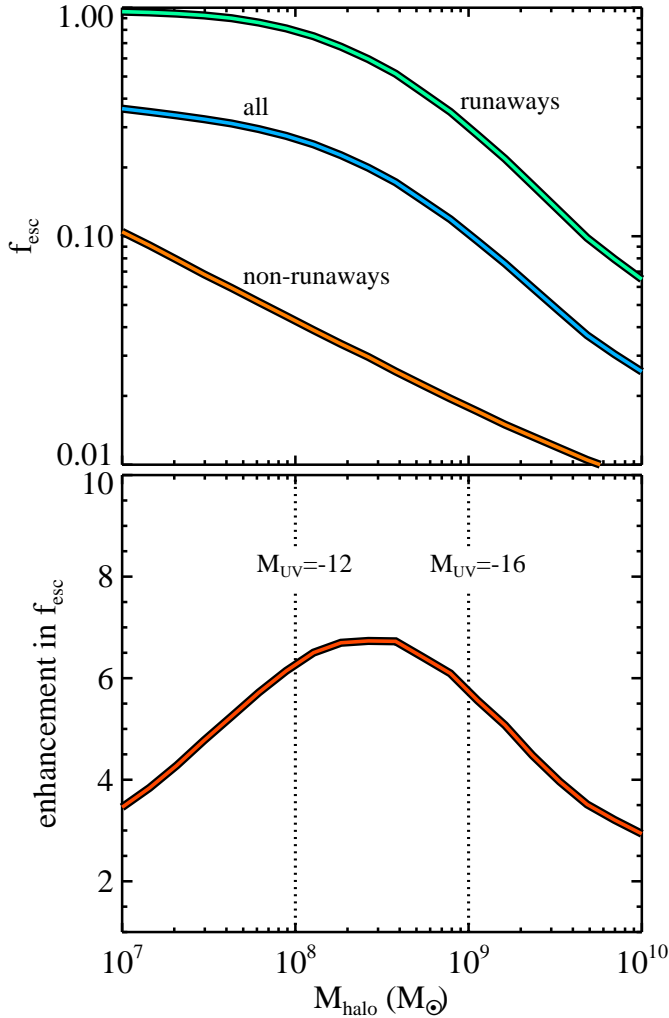}}
\caption{{\it Top Panel:} Escape fraction of ionizing radiation,
  $\fesc$, as a function of halo mass, $M_{\rm halo}$, for our
  fiducial set of model parameters.  The escape fraction is shown for
  all stars, runaways only, and non-runaways only.  {\it Bottom
    Panel:} The enhancement in $\fesc$, defined as the ratio between
  the escape fractions computed with all stars and with non-runaways
  only.  Vertical lines represent the approximate halo masses of
  galaxies with an absolute UV magnitude of -12 and -16
  \citep{Kuhlen12}. }
\label{fig:std}
\end{figure}

\begin{figure*}[!t]
\center
\resizebox{7in}{!}{\includegraphics{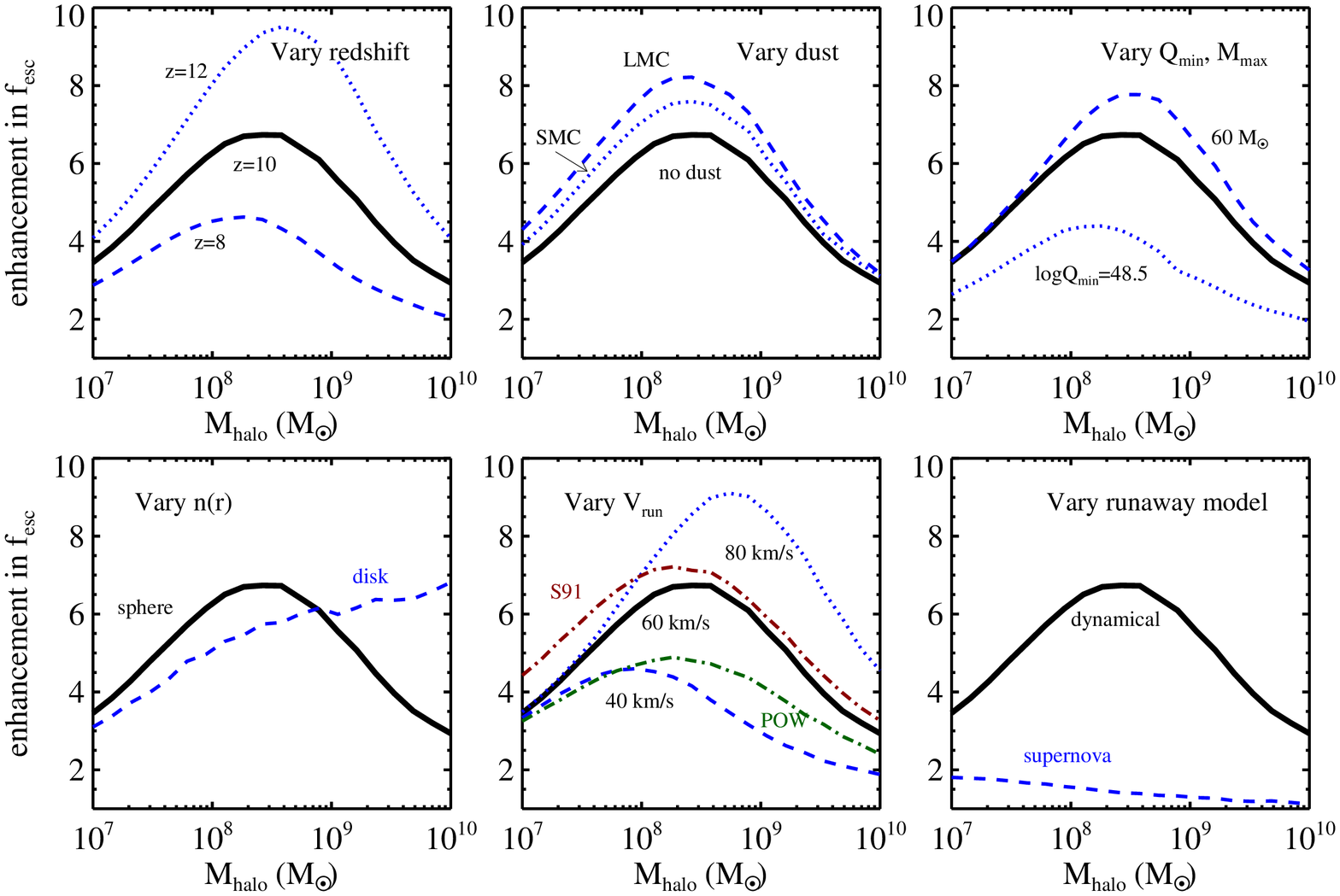}}
\caption{The effect of various model parameters and assumptions on the
  derived enhancement in $\fesc$ due to runaway stars.  The solid line
  in each panel is our fiducial model where runaways are produced by
  dynamical encounters.  {\it Top Left:} Variation in the redshift of
  the model from $8<z<12$.  {\it Top Middle:} Variation in the
  treatment of dust.  {\it Top Right:} Variation in the maximum
  stellar masses of the runaways and in the minimum ionizing
  luminosity, $Q_{\rm min}$, of the combined runaway and non-runaway
  population. In the fiducial model, $M_{\rm max}=100\Msun$ and
  log$(Q_{\rm min})=48.0$. {\it Bottom Left:} Variation in the gas
  density profile.  The fiducial model of a spherical exponential
  profile is contrasted with an exponential disk with a Gaussian
  vertical density distribution.  The disk model has a scale
  height-to-scale length ratio of 0.2 at $\Mhalo=10^8\Msun$ and varies
  as $M^{-2/3}$.  {\it Bottom Middle:} Variation in the treatment of
  the runaway velocities, $\vrun$.  A fixed $\vrun$ is varied from
  $40-80\kms$.  In addition, we consider a power-law distribution of
  $\vrun$ (labelled ``POW'' in the figure), and a Maxwellian
  distribution of $\vrun$ as advocated in \citet[][labelled ``S91'' in
  the figure]{Stone91}.  In this last case $\frun=0.46$, as advocated
  by Stone.  {\it Bottom Right:} Variation in the runaway model
  between our fiducial dynamical model and the supernova model.  See
  the text for details.}
\label{fig:sys}
\vspace{0.1cm}
\end{figure*}

Using the models described above we calculate both the distribution of
$\fesc$ values for massive stars in a given galaxy as well as the
galaxy-averaged $\fesc$ for a range of halo masses.  We begin with the
results from our fiducial model, and then explore the dependencies of
$\fesc$ on a variety of parameter choices.

\vspace{1cm}

\subsection{Fiducial Model}

We show the distribution of $\fesc$ values for stars in our fiducial
model in Figure \ref{fig:pfesc} for both runaways and non-runaways at
a halo mass of $10^8\Msun$.  The vast majority of non-runaways have
$\fesc=0.0$, with only a few percent in the ``translucent'' regime
where $0<\fesc<1$ \citep[see also][]{Gnedin08}.  In contrast, a
significant fraction of runaways have very high $\fesc$, with
$\approx50\%$ having $\fesc>0.6$.  No stars have $\fesc=1.0$.  This is
because stars are never infinitely far from the galaxy, so the galaxy
always occupies a non-zero solid angle capable of absorbing some
ionizing radiation.

Figure \ref{fig:std} contains the main result of this paper.  The top
panel shows $\fesc$ as a function of $\Mhalo$ both for the galaxy as a
whole and separately for the runaways and non-runaways.  The escape
fraction for non-runaways is $<3\%$ for $\Mhalo>10^8\Msun$, in
agreement with previous analytic models of galaxies at $z=10$
\citep{Wood00}.  For non-runaways, $\fesc$ increases toward lower
masses because the ratio of surface area to volume increases toward
smaller galaxies ($\fesc\propto r_g^{-1}$ for non-runaways).  The
escape fraction for runaways, however, is always larger than for
non-runaways, and approaches unity for the smallest galaxies (recall
that $r_g\propto \Mhalo^{1/3}$, making the distance traveled by
runaways in low mass halos a larger fraction of the galaxy size).

The bottom panel in Figure \ref{fig:std} shows the ratio between the
overall escape fraction and the non-runaway value, which we call the
enhancement factor in $\fesc$.  The turnover at $\Mhalo>10^8\Msun$ is
due to the fact that runaways travel on average a fixed physical
distance that becomes a smaller fraction of the galaxy scale radius at
larger masses.  At sufficiently large scale radii the runaway escape
fraction should converge to the non-runaway value, as the runaways
will travel an insignificant fraction of a scale length.  The runaway
$\fesc$ must therefore decrease more rapidly than the non-runaway
value.  The enhancement factor has a maximum because $\fesc$ for
runaways begins to saturate near $\Mhalo=10^8\Msun$.  At lower masses
$\fesc$ for non-runaways continues to increase, so the ratio of
$\fesc$ between all stars and non-runaways decreases at
$\Mhalo<10^8\Msun$.

In this figure we also label approximate UV magnitudes of galaxies in
halos of masses $10^8\Msun$ and $10^9\Msun$ at $z=10$ based on
estimates of the $z=10$ UV LF \citep{Kuhlen12}.  These estimates are
highly uncertain; they are included here merely to indicate the types
of galaxies one might expect to occupy these halos at high redshift.

The key result is that $\fesc$ increases by factors of $2-8$ for
$10^7\Msun<\Mhalo<10^{10}\Msun$ when runaways are included in the
model.  Figure \ref{fig:std} demonstrates that, in our model, runaways
contribute $50-90\%$ of all the ionizing photons that escape from the
galaxy.  Runaways may therefore play an important role in reionizing
the universe.

\begin{figure}[!t]
\center
\resizebox{3.5in}{!}{\includegraphics{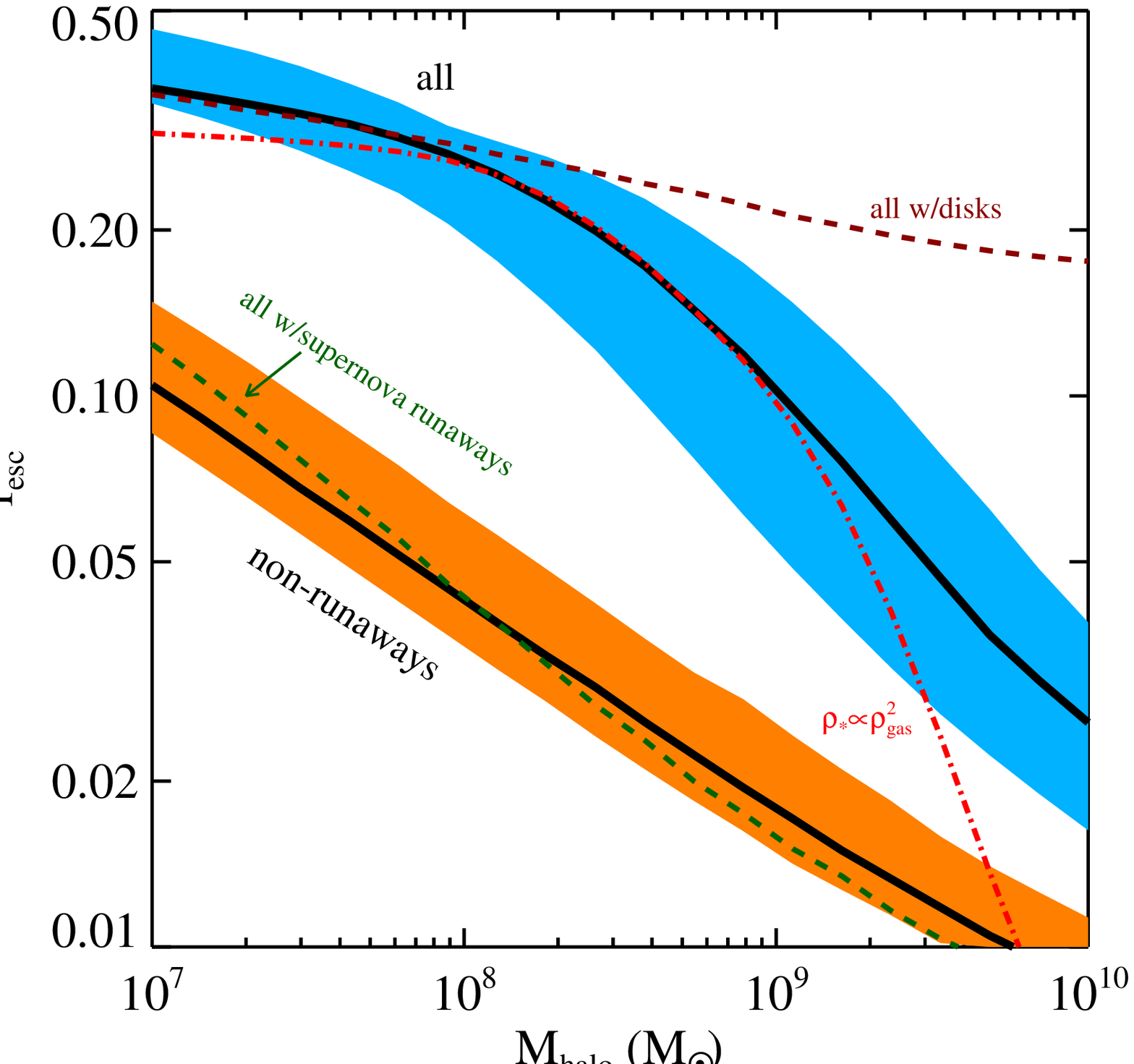}}
\vspace{0.1cm}
\caption{Variation in $\fesc$ between all models considered in the
  text.  The shaded regions encompass the maximum range of the models
  both with and without the inclusion of runaways (labeled ``all'' and
  ``non-runaways'' in the figure).  Thick black lines represent the
  fiducial model.  We have separately plotted the models that deviate
  substantially from the general trend, including models with a disk
  geometry, with supernova runaways, and with the stellar density
  profile scaling as the gas density squared. }
\label{fig:af}
\end{figure}

\subsection{Model Dependencies}

The sensitivity of these results to the assumptions of the model is
explored in Figure \ref{fig:sys}.  In this figure we vary the adopted
redshift from $8<z<12$, the dust attenuation, the minimum and maximum
stellar mass, the adopted gas density profile, $n(r)$, $\vrun$, and
the ionizing LF. We also explore the predictions of the supernova
mechanism for the production of runaways.

The trends are straightforward to interpret. At higher redshift,
galaxies are smaller at fixed mass.  As the size of the galaxy
decreases the effect of runaways becomes stronger both because it is
easier for runaways to move to regions of low column density and
because, at fixed galaxy mass, smaller sizes result in denser
galaxies, lowering the escape fraction for non-runaways
\citep{Wood00}.  For example, between $z=8$ and $z=12$ $\fesc$
increases by 70\% at $\Mhalo=10^9\Msun$. 

Increasing $\vrun$ also results in runaways having a greater effect on
$\fesc$.  In this figure we have also considered distributions of
$\vrun$, rather than single values.  We have included both a power-law
distribution with an index of $-1$ and a mean of $40\kms$, and a
Maxwellian with a dispersion of $30\kms$.  In the latter model we have
simultaneously set $\frun=0.46$ in order to mimic the observational
constraints on the runaway population determined by \citet{Stone91}.
Adopting a distribution of $\vrun$ rather than a fixed value has a
modest effect on the results.

Dust has a very minor effect on the derived $\fesc$ values.  As
pointed out by \citet{Gnedin08}, even without dust the vast majority
of non-runaways already have $\fesc=0.0$ (see Figure \ref{fig:pfesc}),
and so the addition of dust attenuation provides a minor modulation to
the source-averaged $\fesc$.  In addition, the majority of runaways
reside in regions of very low column density, where dust attenuation
also has a minor effect.

The adopted gas density profile has a dramatic effect on the resulting
enhancement factor.  In Figure \ref{fig:sys} we show results for both
a spherical exponential distribution and an exponential disk with a
Gaussian vertical distribution.  Recall that in our disk model, the
scale height-to-scale length ratio varies as $M^{-2/3}$, as expected
for an isothermal disk, with a value of 0.2 at $\Mhalo=10^8\Msun$. In
the case of disks, the runaway escape fractions are very high at high
masses ($\fesc\gtrsim50$\%) because the disks are thin and runaways
can easily escape in the direction perpendicular to the disk.  At
fixed mass, the enhancement in $\fesc$ is a strong function of
$z_g/r_g$, with smaller values resulting in larger enhancements.  At
small masses the scale height-to-scale length ratio approaches unity
in our model, and so the disks are essentially spheres. Thus the
enhancement factor is similar between the two geometries\footnote{The
  enhancement factors are not identical even at the lowest masses
  because a sphere is not identical to a disk with
  $z_g/r_g\approx1$.}.

Reducing the maximum stellar mass of runaways from $100\Msun$ to
$60\Msun$ causes a modest increase in the runaway enhancement because
the mean stellar lifetime of runaways increases as the mass decreases.
The runaways are therefore capable of traveling further from the
central regions of the galaxy before they explode.  We have assumed
here that the runaway fraction does not depend on stellar mass at
$>60\Msun$.  This assumption is supported by the results of
\citet{Stone91}, who finds no variation in $\frun$ amongst the O-type
stars.  When the minimum $Q$ for both the runaways and non-runaways
increases from $48.0$ to $48.5$, the decrease in the mean lifetime of
the runaways results in a lower enhancement in $\fesc$.

The bottom right panel of Figure \ref{fig:std} shows the effect of
adopting the supernova model for runaways rather than the dynamical
mechanism.  Supernova-produced runaways result in a very modest
enhancement in $\fesc$.  The primary difference between these
mechanisms, as implemented herein, is the different lifetime of the
runaways.  Dynamically-created runaways travel for roughly three times
longer than supernova-created runaways, and so the former can more
easily reach lower column densities in the galaxy.  Observations favor
roughly equal contributions from dynamically-produced and
supernova-produced runaways, and so we can expect reality to lie in
between these two extremes. Various stellar evolution models that
predict different lifetimes for massive stars should change $\fesc$ by
a more modest factor.

There are additional parameters, not explicitly shown in Figure
\ref{fig:std}, that can also have a large impact on the enhancement in
$\fesc$.  The runaway fraction, $\frun$, clearly will have a direct
(linear) effect on $\fesc$.  The galaxy size also has a strong effect
on the results, as can be inferred from the upper left panel of Figure
\ref{fig:std}.  In this panel we consider variation in the adopted
redshift, but this is simply changing the galaxy size at fixed mass
(see Equation \ref{e:r}).  The variation in redshift amounts to only a
$\pm20\%$ change in the typical size of a galaxy at fixed mass,
implying that the results do depend sensitively on size.  This
highlights the need for incorporation of runaways into realistic
simulations of high-redshift galaxy formation to assess their effect
on $\fesc$ in a more quantitative manner.  

In Figure \ref{fig:af} we show $\fesc$ as a function of $\Mhalo$ for
the full range of models discussed in this section.  The shaded
regions encompass the total variation in $\fesc$ due to the different
model assumptions.  Results are shown both with and without the
inclusion of runaways.  In the case where runaways are included, we
separately highlight the most deviant models.  These models are the
disk geometry model and the supernova mechanism for the creation of
runaways.  In the disk model, $\fesc$ is much higher than the fiducial
model at high masses because runaways can more easily escape the
galaxy when traveling perpendicular to the disk.  In the supernova
runaway model $\fesc$ is considerably smaller because runaways travel
for a relatively short time before they explode.  In all cases where
runaways are produced via dynamical encounters the escape fraction can
be quite high even in moderately large halos.

Finally, in Figure \ref{fig:af} we separately show a model where the
stellar distribution is proportional to the gas density squared, in
contrast to our fiducial model where the stellar density is linearly
proportional to the gas density.  This model takes into account that
stars form preferentially at the highest densities.  In this model the
escape fraction for non-runaways is $10-20$ times lower than our
fiducial model, as the stars are now much more embedded on average. On
the contrary, the escape fraction of runaways is identical to the
fiducial model for $\Mhalo\lesssim10^9\Msun$.  In larger halos the
galaxy size becomes comparable to, and eventually larger than, the
mean distance traveled by runaways, and so the birth environment of
the runaways becomes increasingly important in determining $\fesc$.
Since the non-runaway $\fesc$ is so much smaller than the runaway
$\fesc$ in this model, the behavior of the runaways entirely
determines the behavior of the overall escape fraction.  As is clear
from Figure \ref{fig:af}, this model produces very similar overall
$\fesc$ values for $M\lesssim10^9\Msun$ compared to the fiducial
model.  At higher masses this model produces lower escape fractions
because the runaways are closer to their birth environment (in units
of scale lengths), and born in denser regions on average when
$\rho_*\propto \rho_g^2$.

\section{Discussion \& Conclusions}

In this paper we have shown that the ionizing radiation from runaway
stars may contribute substantially to the reionization of the
universe. These stars migrate toward the low-density outer regions of
high-redshift galaxies where their radiation can easily escape into
the IGM.  The importance of their migration is enhanced at high
redshift because the galaxies are much smaller than at $z=0$ (by a
factor of $\sim(1+z)^{-1}$).  Assuming that runaways have a prevalence
that is similar to what is observed in the Galaxy ($\sim30\%$ for
massive stars) and that dynamically-created runaways constitute a
significant fraction of all runaways, they can increase the total
escape fraction of ionizing photons from high-redshift galaxies by
factors of $2-8$ compared to the escape fraction of non-runaway stars.

Our conclusions depend strongly on three model ingredients: the size
of the galaxy, the runaway fraction, $\frun$, and the production
mechanism for runaways.  The effect of runaways on $\fesc$ depends
linearly on $\frun$, and therefore if the fraction of runaways is
substantially smaller than what we have assumed here, their importance
in an extragalactic context will be limited.  We have also assumed
that galaxy sizes at high redshift scale with halo size in a manner
similar to what is found at $z=0$, and so high-redshift galaxies are
assumed to be much smaller than local galaxies. Small galaxies
strongly enhance the effect of runaways on $\fesc$.  Finally, we have
shown that runaways produced via dynamical encounters have a much
larger effect on $\fesc$ than runaways produced via an explosion of a
close companion because of the different runaway lifetimes implied by
these models.  Observations of runaways in the Galaxy favor a mixture
of these two mechanisms \citep{Tetzlaff11}, and so we can expect that
runaways in high-redshift galaxies will play an important role in the
escape of ionizing radiation.

The relative importance of runaways depends on the escape fraction of
non-runaway stars.  In our model the non-runaways have $\fesc<10\%$,
and so runaways can be very influential.  However, if the non-runaway
escape fraction were much higher, then the runaways would necessarily
have a diminished impact.  As mentioned in the Introduction, some
recent hydrodynamic simulations have found very high escape fractions
in high-redshift galaxies \citep{Wise09, Razoumov10, Yajima11}.  These
models invoke very efficient feedback that leads to a highly porous
ISM which in turn allows for many unobscured sight-lines to massive
stars \citep[see also][]{Wood00, Fernandez11}.  If these numerical
models are correct, then the influence of runaways on $\fesc$ will be
significantly decreased. To better understand the influence of
runaways on realistic galaxies, they must be included
self-consistently in detailed hydrodynamic simulations that include
the effects of radiative transfer.

The influence of runaways on $\fesc$ can also be tested directly with
observations of high-redshift galaxies.  If the escape fraction of
non-runaways is indeed low, then there will be significant luminosity
from recombination emission lines associated with the gaseous disk.
Beyond a few disk scale lengths the runaways will dominate the stellar
emission, which implies a decrease in the ratio of recombination line
flux to ultraviolet flux.  In other words, the scale length of a
galaxy measured in recombination lines should be smaller than the
scale length measured in broadband ultraviolet flux.  Such an
observation would provide evidence for the influence of runaways
\citep[although we note that other explanations for a varying
recombination line-to-ultraviolet flux ratio have been proposed, see
e.g.,][]{JLee09}.  The ratio of scale lengths of the runaways to
non-runaways is a decreasing function of halo mass: at
$\Mhalo=10^8\Msun$ the ratio of scale lengths is 2.6 while at
$\Mhalo=10^9\Msun$ the ratio is 1.5, for our fiducial model. Such
measurements should be within reach of the next generation of
thirty-meter telescopes.  In addition, the occurrence of supernovae
far from the inner regions of high-redshift galaxies should be common,
and would provide another diagnostic of the prevalence of runaways
during the reionization epoch.

Runaways may have other important effects on the evolution of galaxies
and the IGM.  \citet{Ceverino09} included runaways in a
high-resolution simulation of a Milky Way-like disk galaxy and found
that they lead to much more efficient heating of the ISM.  At high
redshift, runaways will also be able to effectively heat and enrich
the outer regions of halos and the IGM. Binary formation among the
first stars may also have an important effect on IGM heating and
chemical enrichment via the production of Population III runaways.

In this paper we have offered a plausible mechanism for the efficient
escape of ionizing radiation from galaxies during the reionization
epoch.  Runaways almost certainly exist at high redshift, and their
presence will result in an increase in $\fesc$.  Detailed simulations
that incorporate runaways are required in order to make more
quantitative statements regarding their potentially fundamental role
in the reionization of the universe.  Our results also highlight the
need for more detailed observations of the runaway population in the
local universe, including the dependence of $\frun$ on stellar mass,
the runaway velocity distribution, and the relative contribution of
dynamically-produced and supernova-produced runaways.  Fortunately, a
wealth of new data on runaway stars should be available in the coming
years \citep[e.g.,][]{Oey11, Lamb11}.

\acknowledgments 

We thank Andrey Kravtsov, Avi Loeb, and Sally Oey for
comments on an earlier draft. We also thank the referee for a careful
review of this manuscript.


\end{document}